\documentclass[a4paper,11pt]{article}
\usepackage{pos}

\title{A new era for multi-wavelength studies of blazars with Rubin and the CTAO}

\author*[a]{Julian Hamo}
\author[a,b]{Jonathan Biteau}
\author[a]{Julien Peloton}

\affiliation[a]{Université Paris-Saclay, CNRS/IN2P3, IJCLab,\\
  91405 Orsay, France}

\affiliation[b]{Institut Universitaire de France (IUF),\\
France}

\emailAdd{julian.hamo@ijclab.in2p3.fr}
\emailAdd{jonathan.biteau@ijclab.in2p3.fr}
\emailAdd{julien.peloton@ijclab.in2p3.fr}

\abstract{
Both the Rubin Observatory and the first telescopes of the CTAO will be collecting data by 2026, marking a new era in optical and gamma-ray astronomy. Compared to predecessors like the ZTF, H.E.S.S., MAGIC, and VERITAS, their enhanced sensitivity will extend extragalactic observations to a redshift of at least $\sim$2.5. This advancement offers fresh insights into non-thermal astrophysical sources, particularly blazars - radio-loud Active Galactic Nuclei with jets aligned with our line of sight. The 3-night cadence monitoring with Rubin, in one of its six filters, will produce blazar light curves that, when combined with targeted in-depth observations from the CTAO, could help distinguish acceleration and radiative models, which are still under debate. Existing data from the ZTF and \textit{Fermi}-LAT, though less sensitive, offer preliminary insights into what Rubin and the CTAO may achieve. However, the real-time processing of the immense data stream coming from Rubin/LSST presents a major challenge.

Addressing this challenge is the work of brokers such as Fink, which we develop for multi-messenger astrophysics. Fink processes data in real-time before sending relevant information to other observatories like the CTAO. In this contribution, we present how we characterize the optical variability of blazars that emit in the gamma-ray range using the ZTF, with timescales spanning from the intra-night to multi-years. We identify properties in the resulting parameter space that could not only enable the identification of blazar-like sources, but also the characterization of the continuum of states. We describe our fast identification of transitions from one state to another, enabling the trigger of observations in the gamma-ray band when the blazar is flaring and of spectroscopic observations with the goal to measure the redshift of the source when the jet becomes faint and the host galaxy may become detectable. Finally, we review the communication channel we set from the ZTF to the CTAO via Fink for blazars and discuss its outlook in light of the Rubin Observatory. This method is also applicable to other astrophysical sources and helps lay the groundwork for a fruitful era for time-domain astronomy.
}

\ConferenceLogo{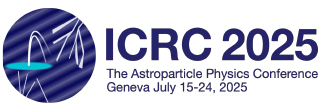}

\FullConference{39th International Cosmic Ray Conference (ICRC2025)\\
 15–24 July 2025\\
Geneva, Switzerland\\}

\begin{document}
\maketitle

\section{Introduction}

Blazars are radio-loud active galactic nuclei (AGNs) with a prominent jet pointing towards our line of sight ($\theta < 10^{\circ}$). Blazars are the main contributors \cite{Urry1995} to the high- and very-high-energy extragalactic $\gamma$-ray sky, via their production of highly relativistic particles. However, their emission and acceleration mechanisms are not yet fully understood. As a result of this acceleration, blazars radiate across the entire electromagnetic spectrum \cite{Ghisellini2017}. Moreover, one of their defining characteristics is their temporal flux variability at all wavelengths \cite{Gaur2012, Aharonian2007}, with power distributed over all timescales. These properties make blazars difficult sources to model, as different processes may be involved in different emission states. This is particularly the case for flaring states. Such flares are prominent in both the optical-to-soft X-ray and hard X-ray-to-$\gamma$-ray ranges, hence the interest in correlation studies.

To observe different states of a blazar simultaneously in the optical and gamma-ray bands, continuous monitoring is required — an ideal role for optical large-scale surveys like the Zwicky Transient Factory (ZTF) and the Legacy Survey of Space and Time (LSST). The ZTF, operating since 2018 from the Palomar Observatory (California, USA), covers the northern hemisphere down to –30$^{\circ}$ declination with a cadence of 2 days and a depth of magnitude 20.4 using three filters: g (478 nm), r (642 nm), and i (787 nm) \cite{Bellm2019}. Its successor, the LSST survey, begins mid-2025 at the Vera C. Rubin Observatory (Chile), covering the southern hemisphere up to +12$^{\circ}$ declination every 3 days, reaching magnitude 23, and using six filters: u, g, r, i, z, and y (368, 478, 622, 753, 869 and 973 nm) \cite{Ivezic2009}. Both surveys yield dense but unevenly sampled multi-band light curves.

On the other hand, $\gamma$-ray light curves are produced by the \textit{Fermi} Large Area Telescope (LAT) \cite{Atwood2009}. Launched in 2008, the LAT observes the sky in the 100 MeV-1 TeV energy range. However, due to its orbit and small size, the \textit{Fermi}-LAT does not permit deep observations of a source at a specific time. This task is better suited to the forthcoming Cherenkov Telescope Array Observatory (CTAO), located on the island of La Palma in the Canary Islands for the northern site and in Chile for the southern site. The energy range from 0.02 to 200 TeV, energy resolution of <10\% and km$^2$ effective area of the CTAO allow for better spectral reconstruction and time sampling for a specific observation than the \textit{Fermi}-LAT \cite{Hofmann2023}.

Using the LSST and the CTAO together would enable precise multi-wavelength measurements of the various states of blazars to be taken. However, a communication channel is required between the two observatories to handle the important data stream from the LSST (10M alerts per night), select the relevant alerts, and send formatted trigger proposals to the CTAO. Such a channel is made possible thanks to brokers such as Fink \cite{Moller2020}. This broker has been retrieving g- and r-band alerts from the ZTF in real time since 2019. Fink is also connected to other observatories, such as the Siding Spring Observatory, as well as to other messengers via GCN, making it an ideal tool for multi-messenger astronomy.

\section{Dataset}

The 3rd \textit{Fermi}-LAT Catalogue of High-Energy Sources 3FHL \cite{Ajello2017} includes all sources significantly detected by the \textit{Fermi}-LAT in the 10 GeV-2 TeV energy range during its first 7 years of operation. 
We retrieve blazars from this catalogue, which are classified as flat-spectrum radio quasars (FSRQs), BL Lacs (BLLs) and blazar candidates of an uncertain type (BCUs). There are 1212 blazars in the 3FHL: 172 FSRQs, 750 BLLs and 290 BCUs.

\begin{figure*}[t]
    \centering
    \includegraphics[width=\textwidth]{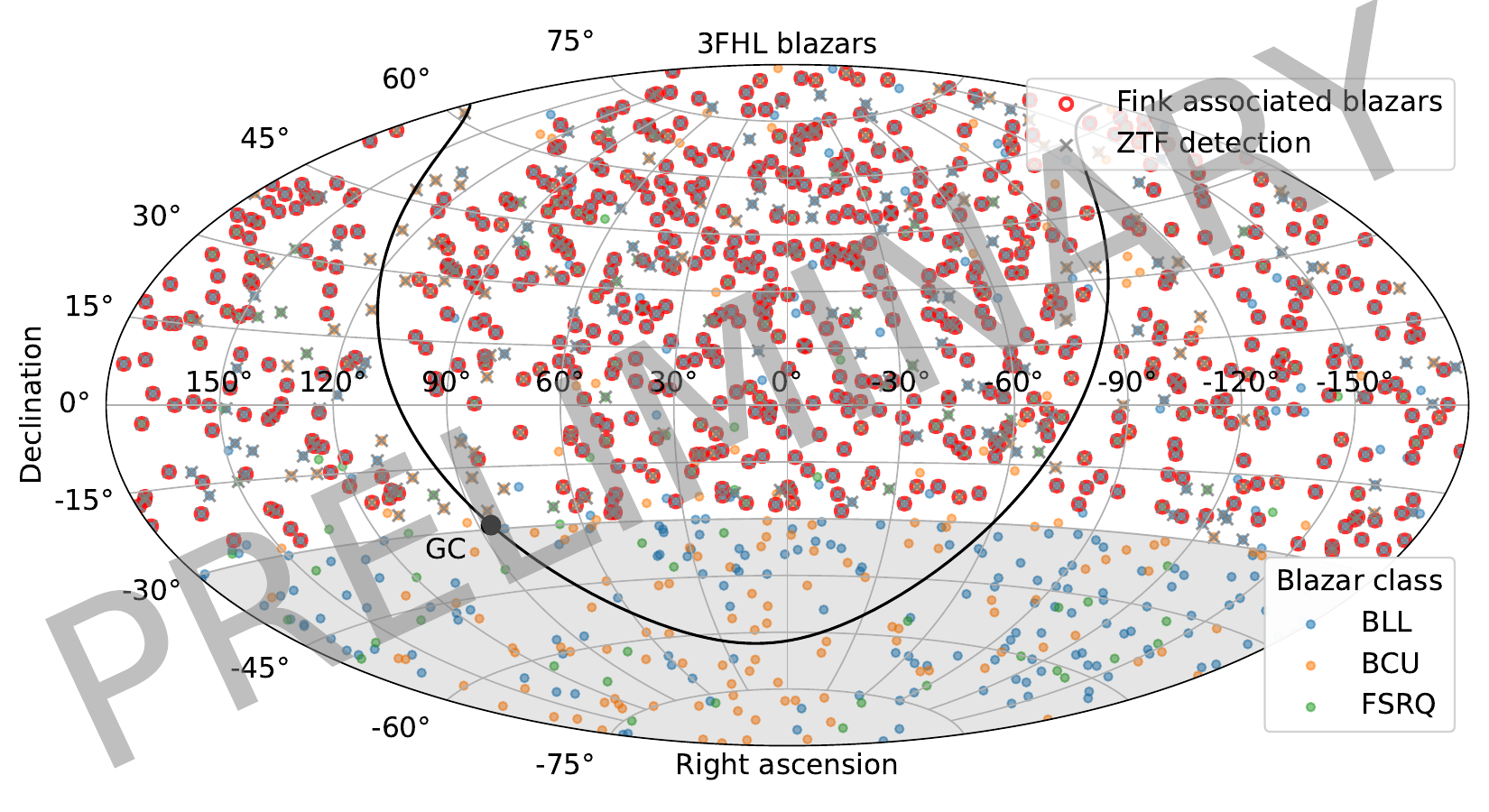}
    \caption{Skymap of the 3FHL blazars (BLLs in blue, BCUs in orange and FSRQs in green) in equatorial coordinates. The grey crosses show the 807 sources that have been detected by the ZTF in these directions. The red circles indicate the 621 ZTF sources that are classified as blazars by Fink. The greyed area is the part of the sky that is not visible to the ZTF. The Galactic Plane and the Galactic Centre (GC) are shown as a black line and a black dot, respectively.}
    \label{fig:skymap}
\end{figure*}

We query their most accurate coordinates via the Simbad API \cite{Wenger2000} and perform a cone search in the Fink database within a radius of 1 arcminute. We use the classification modules of Fink to associate each source with the closest ZTF association candidate that has been classified as a possible blazar by Fink.
Only 621 3FHL blazars are associated in Fink (67\% of matches for blazars that are in the part of the sky covered by the ZTF). The fraction of matches rises to 87\% when considering the ZTF data release \cite{Masci2019}. The missing sources can be explained by a too crowded forefront in the Galaxy. This suggests possible improvements to the Fink classification algorithms, particularly along the Galactic Plane, as illustrated in Fig~\ref{fig:skymap}.

\section{Identification of extreme states}

Although blazar variability is an extensively studied topic, there is no consensus on how to define an extreme state. We develop two approaches to identify these extreme states, whether they are high or low states.

\subsection{Standardisation process}

If two light curves from the same source at different wavelengths are correlated, variations in one can also be observed in the other. Merging the two light curves then doubles the amount of data without altering the overall variation behaviour. However, the absolute flux values depend on the observed wavelength. One approach to merging the light curves is to scale them by dividing them by their median flux, a quantile that is independent from the tracer (flux or magnitude), unlike the mean. In the following, this process is referred to as standardisation of the light curves.

\begin{figure*}[t]
    \centering
    \includegraphics[width=\textwidth]{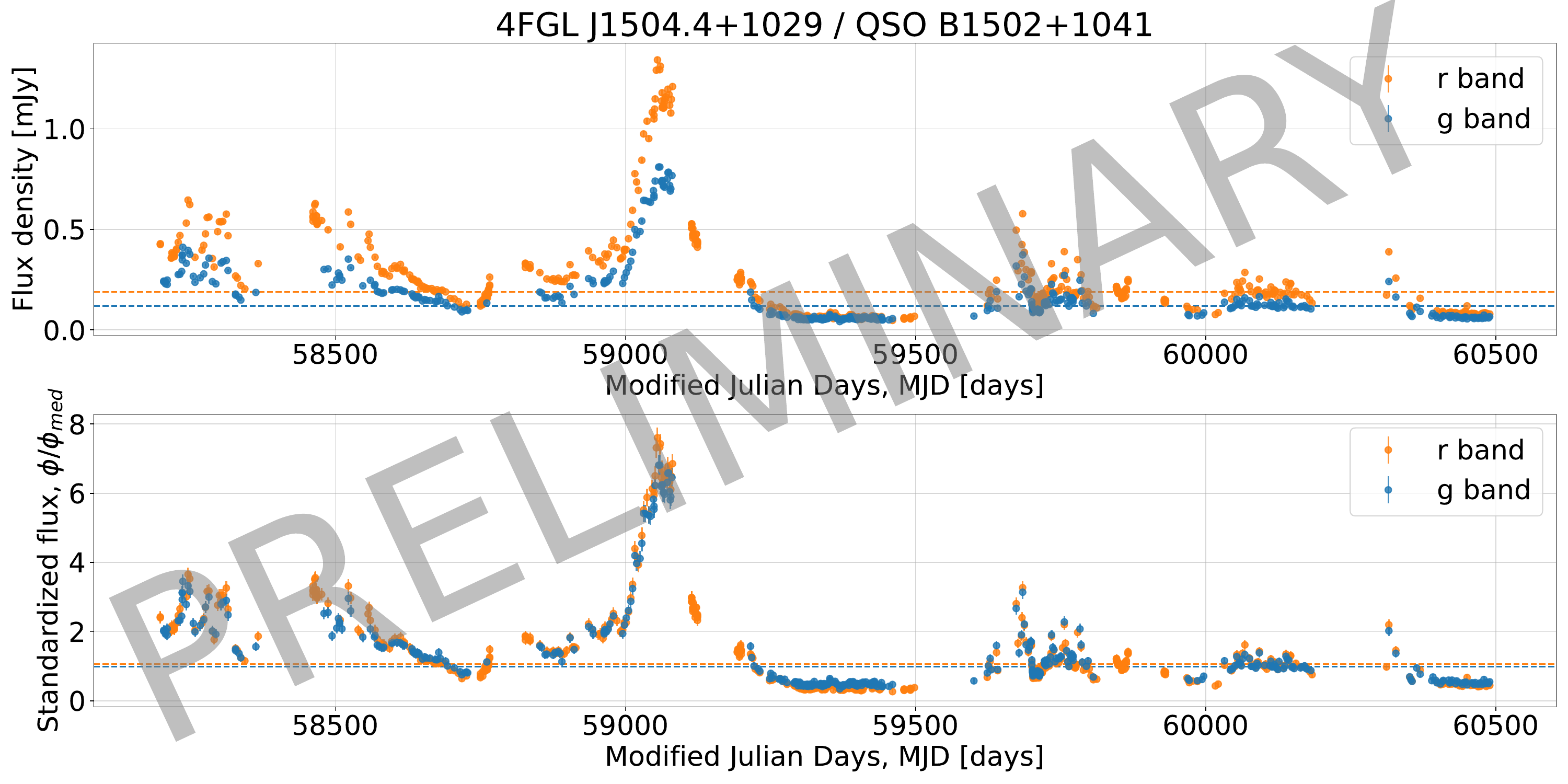}
    \caption{Top: Flux density in milli Jansky, in the r-band (in orange) and g-band (in blue) of the illustrative source 4FGL\,J1504.4+1029, as observed by the ZTF between March 2018 and June 2024. Bottom: Standardised flux computed by dividing the flux by the median of concomitant measurements in both bands. Standardised light curves share common amplitude. In both panels, the dotted coloured lines represent the median of the band of the same colour.}
    \label{fig:standardization}
\end{figure*}

We use light curves in the r- and g-bands from the latest ZTF survey (data release and direct alerts though the Fink broker). We choose only concomitant measurements from the ZTF light curves to compute their median. In practice, for each measurement in one band, we calculate the average measurement that has been taken in the other band within 12 hours (from $t - 6h$ to $t + 6h$), if possible. We store these subsets of flux measurements and calculate their median in each band. These are then used as reference medians to standardise the light curves, as shown in Fig.~\ref{fig:standardization}. This ensures that the medians are not biased by the different cadences (e.g., extended deep drilling programme of the ZTF) or possible gaps in one of the bands.

\begin{figure*}[t]
    \centering
    \includegraphics[width=\textwidth]{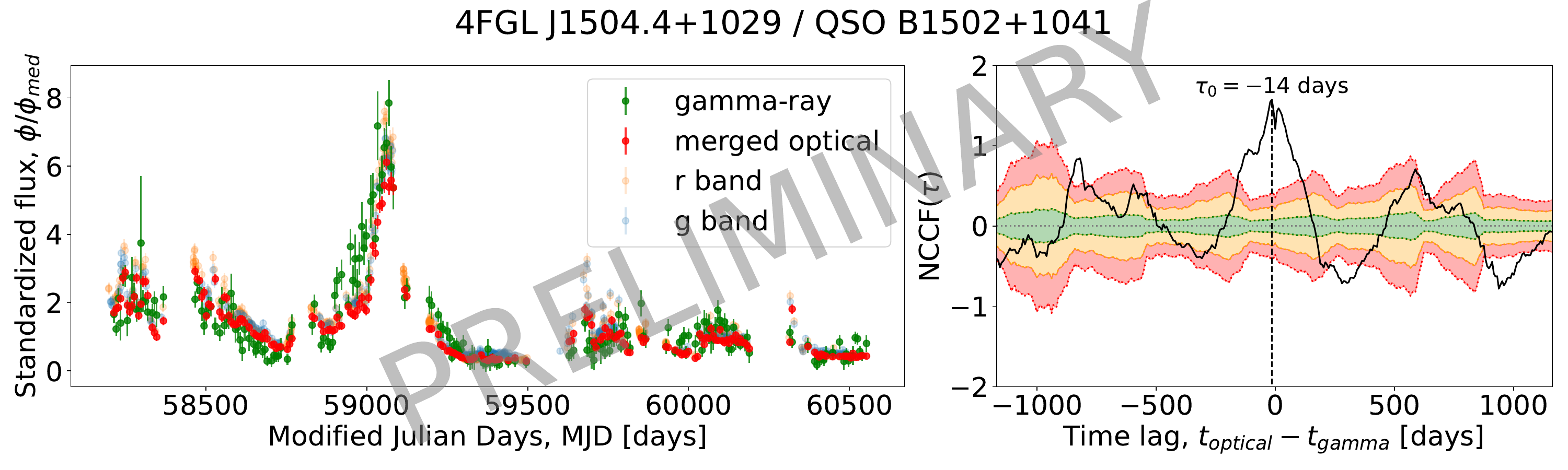}
    \caption{Left: Standardised resampled optical band (in red) and gamma-ray band (in green). Right: Normalised cross-correlation as a function of the time lag. Positive time lag values indicate a $\gamma$-ray light curve that begins after the optical light curve. The edges of the green, orange and red bands represent local p-values of $0.32$, $2.7 \times 10^{-3}$ and $5.7 \times 10^{-7}$, respectively. The dashed vertical line shows the most significant time lag, $\tau = -14 \pm 100$ days, with a significance of 19 $\sigma$.}
    \label{fig:NCCF}
\end{figure*}

We can compare the obtained optical light curve with the $\gamma$-ray flux measurements from the \textit{Fermi}-LAT Light Curve Repository \cite{Abdollahi2023}. As the collection area of the \text{Fermi}-LAT is limited, we choose to study weekly binned $\gamma$-ray light curves to achieve a sufficiently large rate of flux measurements in comparison to upper limits. Similar weekly sampling must be applied to the standardised optical light curve to compare variations on similar timescales. Once applied (Fig.~\ref{fig:NCCF}, left), both light curves can be compared, with the optical light curve being considered as one band and the $\gamma$-ray light curve as another. 

Correlated behaviour suggests a common origin for both phenomena. A statistically different time lag between bands suggests that the emission process responsible for the earlier light curve could help to trigger observations in the other bands. It is therefore useful to compute their (pseudo-)normalized cross-correlation \cite{Edelson1988}, defined as

\begin{equation}
    \label{eq:NCCF}
    \text{NCCF}(\tau) = \frac{1}{N - n_{\tau} + 1} \frac{1}{\sqrt{(\sigma_x^2 - \overline{e}_x^2)(\sigma_y^2 - \overline{e}_y^2})} \sum_{i} (x_{i+n_{\tau}} - \overline{x})(y_i - \overline{y})
\end{equation}

where $\{x_i\}$ and $\{y_i\}$ respectively designate the optical and the gamma-ray bands, of length $N+1$, with respective uncertainties $\{e_{x,i}\}$ and $\{e_{y,i}\}$. $\sigma_x$ and $\sigma_y$ are the standard deviations and $\overline{e}_x^2$ and $\overline{e}_y^2$ are the mean squared uncertainties in the flux. Finally, $n_{\tau}$ is the index representing a time lag of $\tau$ (note that the renormalisation factor differs for unevenly sampled light curves).

The significance of the cross-correlation is also computed using the standard deviation of the function after shuffling the flux values in one band, as shown in Fig.~\ref{fig:NCCF} (right) and calculated in Eq.~(\ref{eq:sigma_NCCF}),

\begin{equation}
    \label{eq:sigma_NCCF}
    \sigma_{\text{NCCF}}(\tau) = \frac{1}{N-n_{\tau}+1} \frac{1}{\sigma_x} \sqrt{ \left( 1 + \frac{1}{N} \right) \sum_{i=0}^N x^2_{i+n} - \frac{1}{N} \left( \sum_{i=0}^N x_{i+n} \right)^2}.
\end{equation}

For the specific example shown in Fig.~\ref{fig:NCCF}, a significant cross-correlation between the $\gamma$-ray and optical light curves is observed, with a short delay (compatible with a zero-time lag) between the two. This hints towards the use of optical flares to trigger $\gamma$-ray observations.

\subsection{Archival light curves}

\begin{figure*}[t]
    \centering
    \includegraphics[width=\textwidth]{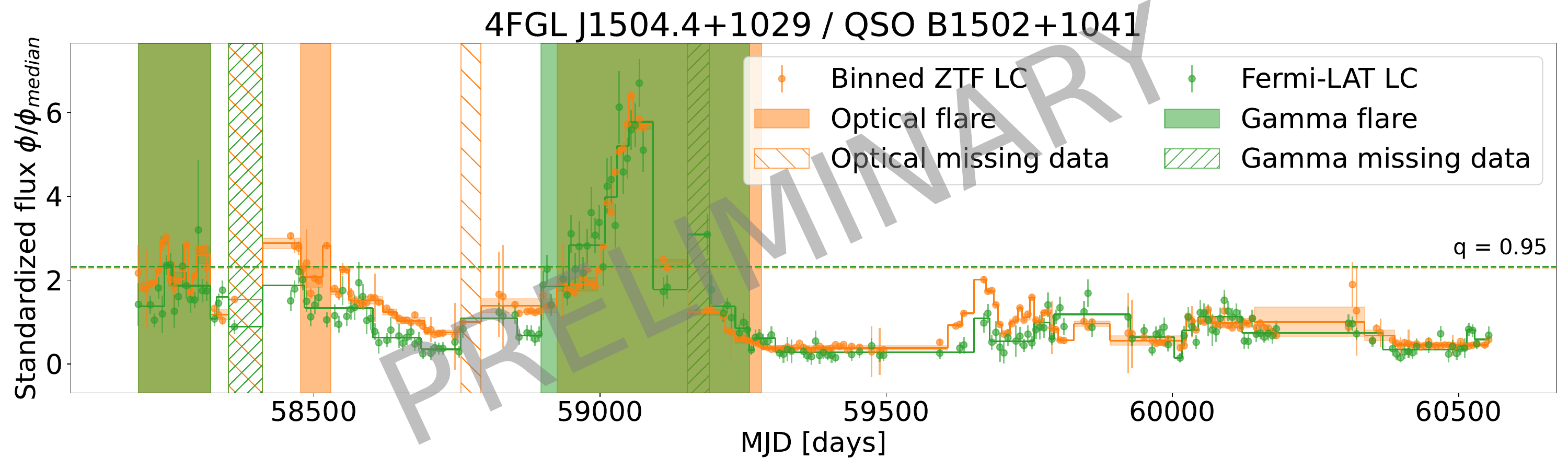}
    \caption{Standardised resampled optical (in orange) and gamma-ray (in green) light curves from the ZTF and \textit{Fermi}-LAT for the illustrative blazar 4FGL\,J1504.4+1029. The solid lines represent the Bayesian block mean fluxes. The dashed horizontal lines show the 95\% quantile for both light curves. The hatched areas show Bayesian blocks with less than 1 point per month and the coloured zones show the high state regions.}
    \label{fig:method1}
\end{figure*}

In the following, we focus on the standardised optical light curves of the blazars under study. We consider a state to be extreme if its flux is above the 90\% quantile (or below the 10\% quantile). Following the method of \cite{Resconi2009}, we first compute the Bayesian blocks \cite{Scargle2013} of the optical light curve to ensure that the crossing of the relevant percentile is not a statistical fluctuation of the emission processes of the blazar but rather a lasting state. We remove Bayesian blocks with insufficient measurements, setting a density threshold of 1 point per month (hashed regions in Fig.~\ref{fig:method1}). 

We identify the Bayesian blocks with a mean above the threshold. For each region above the threshold, we select the Bayesian block with the maximum mean flux. We then merge this block with the adjacent blocks if their mean flux is lower than that of the previous block. We keep repeating this process until the mean fluxes of the considered blocks are no longer lower than those of the previously considered blocks. This process defines a high state. If two high states are consecutive, we merge them. The resulting high states are shown in Fig.~\ref{fig:method1}. A symmetric method is also employed to identify low states.

\subsection{Real-time detection of extreme states}

As seen in Fig.~\ref{fig:NCCF} and Fig.~\ref{fig:method1}, the temporal behaviour of some optical and $\gamma$-ray light curves appears to be correlated. Hence, optical observations can help trigger potentially interesting $\gamma$-ray observations. However, the previously introduced method relies on analysis of the full optical light curve to construct Bayesian blocks. This should be adapted for real-time triggering. ZTF provides Fink with data from the previous 30 days before an alert (the mean duration of Bayesian blocks is about 40 days) and Fink requires alerts to be processed internally within 30 seconds maximum. To address these constraints, we have designed a real-time algorithm to detect extreme states.

We calculate the threshold using the same archival light curves as before. For each new alert (and its 30-day history), we check whether it is above (below) the precomputed corresponding threshold: this ensures that the state being studied is extreme. To ensure that the state is persistent, we calculate the average flux over the history of the alert and check whether it is also greater (or lower) than the threshold. If both the local and the 30-day integrated criteria are met, the alert is considered to be in a robust extreme state. The robustness criterion introduces a typical lag capped to 10 days with a median remaining observation time of 3 weeks.


This low state detection module has already been implemented in Fink for the blazars studied by the CTAO redshift task force \cite{DAmmando2024}, with a median trigger alert rate of $\sim$0.5 per year per source and low states alerts are accessible and regularly sent to CTAO task force.

\section{Outlook: optical characterisation of \texorpdfstring{$\gamma$}--ray blazars}

\begin{figure*}[t]
    \centering
    \includegraphics[width=\textwidth]{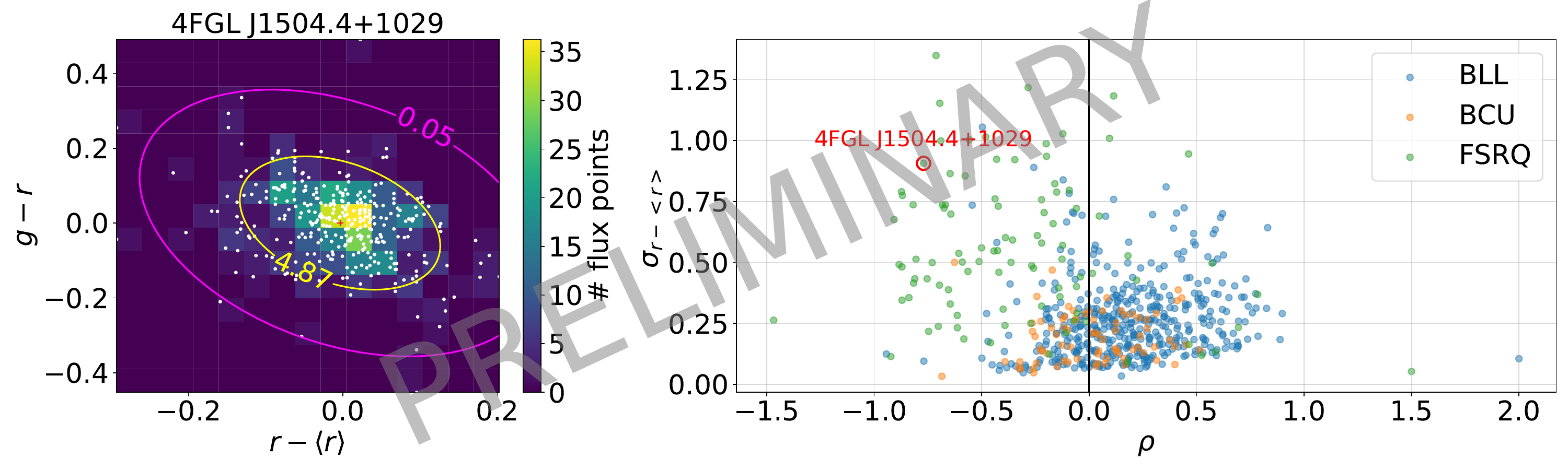}
    \caption{Left: Residual colour of the light curve of the blazar 4FGL\,J1504.4+1029 as a function of the variation of r-band magnitude. The ellipses represent the $1\sigma$ and $2\sigma$ of the 2D Gaussian fit. Right: Scale along the magnitude excursion axis $r - \langle r \rangle$ of the 2D Gaussian fit as a function of the Pearson coefficient $\rho$. BLLs, BCUs and FSRQs are shown in blue, orange and green, respectively. The illustrative source 4FGL\,J1504.4+1029 is circled in red.}
    \label{fig:color_domain}
\end{figure*}

Another application of the multi-wavelength optical light curves from the ZTF is characterising (and potentially identifying) blazars. The standardisation process removes the global colour of a blazar, which can hide relevant information. The authors of \cite{Negi2022} studied the colour variation of blazars in different states. Following this approach, we examine the colour $g-r$, as a function of the variation in the r-band magnitude around the global median $r - \langle r \rangle$, for measurements in both bands no more than 12 hours apart. We fit a 2D Gaussian to the resulting distribution, with the parameters being the 2D mean, the 2D scale and the Pearson correlation coefficient, $\rho$. An example of such a fit is shown in Fig.~\ref{fig:color_domain} (left).

On Fig.~\ref{fig:color_domain} (right), we show the variations of $\rho$ and $\sigma_{r - \langle r \rangle}$ over the 3FHL blazar population. BLLs appear to exhibit lower-amplitude variations away from their median magnitude (with an average value $\sigma_{r - \langle r \rangle}$ of $0.26$) than FSRQs ($\overline{\sigma_{r - \langle r \rangle}} = 0.56$). Moreover, BLLs tend to exhibit a positive Pearson coefficient (325 out of 444 sources, or 73\%) whereas FSRQs tend to exhibit a negative one (81 out of 96 sources, or 84\%). This suggests that BLLs tend to be bluer-when-brighter blazars, whereas FSRQs tend to be redder-when-brighter blazars, as previously suggested in \cite{Negi2022}. Although BCUs do not exhibit a trend in their Pearson coefficient (53\% are positive), their mean distribution, $\overline{\sigma_{r - \langle r \rangle}} = 0.2$ suggests that most of them are BL Lacs.

\section{Summary}

A multi-wavelength comparison of the long-term light curves of a set of 621 blazars reveals interesting correlations between the optical and $\gamma$-ray bands. An analysis of the correlations could help to constrain the emission and acceleration processes at work in blazars. The standardisation process developed in this study enables several light curves from different wavelengths to be analysed simultaneously. The normalised cross-correlation function describes the significance of the correlations. We restricted the study case to $\gamma$-ray emitting blazars from the 3FHL catalogue, though the developed method is designed to be applicable to LSST and CTAO measurements. 

From this new perspective, we arbitrarily defined extreme states (high and low), using archival optical data from the ZTF and $\gamma$-ray data from Fermi-LAT. These extreme states appear to match well in the optical and $\gamma$-ray bands, suggesting a common origin. The $\gamma$-ray study of extreme states using observatories of precision such as the CTAO can be triggered by extreme behaviour in the optical band, as monitored by optical Large Scale Surveys (ZTF and LSST) and conveyed through brokers such as Fink. This highlights the necessity of developing real-time detection algorithms.

The optical light curves of blazars can also be studied independently of their $\gamma$-ray counterparts. We analysed the residual colour of these light curves to classify blazars. The Pearson coefficient between the colour and broad-band flux level may help to discriminate between BLLs and FSRQs, as well as classify BCUs. A more extensive study of the properties of these light curves is ongoing. One additional benefit of such a study could be distinguishing blazars from other sources.

\end{document}